\documentclass[pre,aps,twocolumn,showpacs,showkeys,floatfix]{revtex4-1}
\usepackage{amsmath}
\usepackage{graphicx}

\begin{document}

\title{Simulation of the two-dimensional Potts model using nonextensive statistics}
\author{A. Boer}
\affiliation{``Transilvania'' University of Bra\c sov, Physics Department, B-dul Eroilor 29, 500036, Bra\c sov, Romania}
\email{boera@unitbv.ro}
\date{\today}

\begin{abstract}
The standard Potts model is investigated in the framework of nonextensive statistical mechanics. We performed Monte Carlo simulations on two-dimensional lattices with linear sizes ranging from 16 to 64 using the Metropolis algorithm, where the classical Boltzmann-Gibbs transition probabilities were modified for the nonextensive case. We found that the Potts model undergoes a phase transition in the nonextensive scenario. We established the order of the phase transition and we computed the critical temperature for different values of the Tsallis entropic index. 
\end{abstract}


\maketitle

\section{Introduction} \label{sec:1}
In the last years nonextensive statistical mechanics was successfully applied in the study of various physical systems, both from theoretical and experimental point of view. Regarding magnetic systems these studies refer to the well known Ising model \cite{lima1999monte,crokidakis2009finite}, connected with some experimental evidences of nonextensivity in the case of CMR manganites \cite{reis2002evidences, reis2003magnetic}.

In the present paper we focus on Monte Carlo simulation of the standard Potts model using nonextensive statistics. It is known from classical statistical mechanics that the Potts model has a richer behavior and it includes the Ising model as a particular case. Therefor the study of this model in the framework of Tsallis statistics can be of interest. In the first part of the paper we review some theoretical aspects regarding the Potts model and nonextensive statistical mechanics. Section \ref{sec:MC} deals with the computational methods used for simulations. In Section \ref{sec:results} we will present and discuss the simulation results. Some conclusions and final remarks will be drawn in Section \ref{sec:conclusions}.

\section{The Potts model} \label{sec:Potts}

The standard Potts model is a generalization of the well known Ising model, however it has a much richer phase structure. The two-dimensional $r$-state Potts model is a lattice of spins $s_i$, which can take the following values:
\begin{equation}\label{eq:1}
s_i=1,2,\ldots,r
\end{equation} 
The interaction Hamiltonian is of the form:
\begin{equation}\label{eq:2}
\mathcal{H} = - J \sum_{\langle i,j\rangle} \delta(s_i, s_j)
\end{equation} 
where the sum is performed over all pairs of first neighbours in a square lattice of size $L\times L$. In the above relation $J$ is a coupling constant and $\delta(s_i,s_j)$ is the Kronecker delta:
\begin{equation}\label{eq:3}
\delta(s_i,s_j)=
\begin{cases}
1 & \text{if } s_i=s_j \\
0 & \text{if } s_i\neq s_j
\end{cases}
\end{equation}
The 2-state Potts model is equivalent to the Ising model, the numerical value of the critical temperature in the former being one-half of its value in the latter.

In classical statistical mechanics the critical temperature of the $r$-state Potts model is given by the relation \cite{wu1982potts, wu2009exactly}
\begin{equation}\label{eq:4}
T_c=\frac{1}{\ln(1+\sqrt r)}
\end{equation}
Strictly speaking the two-state Potts model is equivalent to the Ising model only if we consider the following expression for the hamiltonian:
\begin{equation}\label{eq:5}
\mathcal{H} = \sum_{\langle i,j \rangle} \left[ 1 - 2 \delta(s_i,s_j) \right]
\end{equation}
In the framework of nonextensive statistical mechanics the simulation results obtained for the 2-state Potts model are identical with those obtained for the Ising model (see Ref. \cite{crokidakis2009finite}) only if we choose the Hamiltonian given by Eq. \eqref{eq:5}. This case will be discussed in a forthcoming article.

The ferromagnetic version of the Potts model ($J>0$) undergoes a first order phase transition when $r>4$ and a second order phase transition when $r<4$. These results are known from classical (i.e. extensive) statistical mechanics. In the present paper we will investigate the behavior of the standard Potts model described by the Hamiltonian \eqref{eq:2} using nonextensive statistics.

\section{Nonextensive statistical mechanics} \label{sec:Tsallis}
In 1988 C. Tsallis proposed the following formula for entropy, which aims to generalize the classical Boltzmann-Gibbs statistics \cite{tsallis1988possible,tsallis2009introduction}:
\begin{equation}\label{eq:6}
S_q=k \frac{1-\sum_{i=1}^W p_i^q}{q-1}
\end{equation}
In the above relation $p_i$ denotes the probabilities of various microstates, $k$ is a positive constant and $W$ represents the number of microstates compatible with a given macroscopic state of the system. In equation \eqref{eq:6} $q$ is a real parameter, known as the Tsallis entropic index. For $q=1$ we recover the classical Boltzmann-Gibbs statistics.

Nowadays Tsallis' formula is widely used in the study of various physical phenomena. Tsallis statistics became of interest in the study of nanoscopic systems after the equivalence between Hill's nanothermodynamis and Tsallis thermostatistics has been proved \cite{garcia2005correct}. In this context we consider that applying Tsallis statistics for spin lattices and magnetic systems can lead to new developments in the study of phase transitions.

The most important property of the Tsallis entropy is its nonextensive nature. If we consider a system which consists of two subsystems, $A$ and $B$, then the entropy of the global system is given by the relation \cite{tsallis2009introduction}:
\begin{equation}\label{eq:7}
S_q(A+B) = S_q(A) + S_q(B) + \frac{1-q}{k} S_q(A) S_q(B)
\end{equation}
One can observe that for $q=1$ the entropy becomes extensive.

In nonextensive statistics the thermodynamic quantity associated with a random variable $X$ is given by the $q$-mean value of the respective variable:
\begin{equation}\label{eq:8}
\langle X\rangle_q = \frac{ \sum_i p_i^q X_i }{\sum_i p_i^q}
\end{equation}
If we introduce the escort probabilities
\begin{equation}\label{eq:9}
P_i = \frac{p_i^q}{\sum_i p_i^q}
\end{equation}
then the $q$-mean value of $X$ can be written as
\begin{equation}\label{eq:10}
\langle X \rangle_q = \sum_i P_i X_i
\end{equation}
This relation has the same form as the one used in classical statistical mechanics, but the probabilities $p_i$ are replaced with the escort probabilities defined by means of relation \eqref{eq:9}.

The \emph{canonical distribution} in Tsallis statistics has the following form:
\begin{equation}\label{eq:11}
p_i = \frac{1}{Z_q} \left[ 1 - \frac{(1-q)\beta(E_i-U_q)}{\sum_i p_i^q} \right]^{\frac{1}{1-q}}
\end{equation}
where $p_i$ is the probability corresponding to the energy level $E_i$, $\beta$ is the Lagrange multiplier associated with the energy constraint, $Z_q$ is the generalized canonical partition function and $U_q$ represents the generalized internal energy (defined as the $q$-mean value of the energy).

The canonical distribution given by relation \eqref{eq:11} is self-referential, so it is not suitable for simulations based on the Monte Carlo method. To overcome this problem we can use the so-called $\beta-\beta'$ transformation \cite{tsallis1999nonextensive}:
\begin{equation}\label{eq:12}
\beta'= \frac{\beta}{\sum_i p_i^q + (1-q)\beta U_q}
\end{equation} 
Using the above transformation we obtain for the canonical distribution the following expression:
\begin{equation}\label{eq:13}
p_i = \frac{1}{Z'_q} \left[ 1-(1-q)\beta' E_i \right]^{\frac{1}{1-q}}
\end{equation}
This form of the canonical distribution is suitable for Monte Carlo simulations.

\section{Monte Carlo simulation of the Potts model using Tsallis statistics} \label{sec:MC}

Let us consider a transition from a state of energy $E_i$ to another state of energy $E_j$. If we use the escort probabilities, the acceptance ratio of the transition is given by
\begin{equation}\label{eq:14}
P_{i\to j} = \left[ \frac{1-(1-q)\beta' E_j}{1-(1-q)\beta' E_i} \right]^{\frac{q}{1-q}}
\end{equation}
Based on the above relation we can generalize the standard Metropolis algorithm by replacing the Boltzmann-Gibbs acceptance probability with the acceptance ratio given by equation \eqref{eq:14}.

The definition of the physical temperature in Tsallis statistics is still an open problem \cite{abe2001nonextensive,toral2003definition,abe2006temperature}. Based on experimental data in Ref. \cite{reis2002evidences} it was suggested that we can consider $1/\beta'$ as a temperature scale. This approach was used in Ref. \cite{crokidakis2009finite} by Crokidakis et. al. in the simulation of the two-dimensional Ising model. In the present article we will follow the same path.

We performed Monte Carlo simulations on two-dimensional lattices with linear sizes of 16,24,32,48 and 64. For the Tsallis entropic index we considered the following values: 1, 0.95 and 0.9. As mentioned above, we assumed that $1/\beta'$ can be used as a temperature scale. For every value of $\beta'$ we performed $10^6$ equilibration steps and $7.2\cdot 10^7$ production steps. The hardware setup used for the simulations consists of 18 Intel Xeon processor cores. The software was written in the C programming language, based on \textsf{MPICH2} \cite{MPICH2} for multiprocessing capabilities and \textsf{SPRNG} \cite{SPRNG} for random number generation.

\section{Simulation results} \label{sec:results}
\begin{figure*}[t]
\begin{center}
\includegraphics[width=0.45\textwidth]{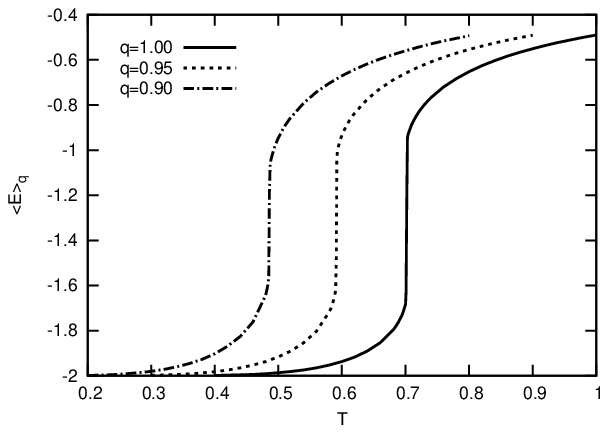} \hspace{0.5cm}
\includegraphics[width=0.45\textwidth]{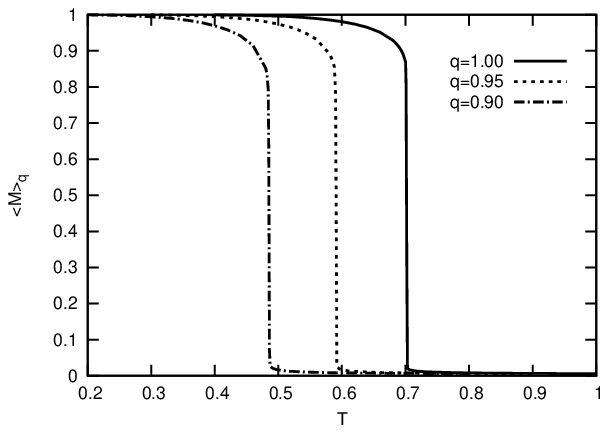}
\end{center}
\caption{Plot of the $q$-mean value of energy and magnetization as function of temperature in the case of the ten-state Potts model, for $L=64$ and different values of $q$.}
\label{fig:1}
\end{figure*}
First of all let us refer to the simulation results obtained for the 10-state Potts model ($r=10$). Figure \ref{fig:1} shows the plot of the $q$-mean value of energy and magnetization as function of temperature for $L=64$ and different values of the Tsallis entropic index. We can observe that the phase transition is present even for $q\neq 1$ and the critical temperature has a strong dependence on $q$.
For the Potts model the magnetization is defined by means of the relation:
\begin{equation}\label{eq:15}
M=\frac{ \sum_{i=1}^N \left[ r \delta(s_i,1) -1 \right] }{N(r-1)}
\end{equation}
where $N$ represents the number of spins in the lattice.

\begin{figure}[h]
\begin{center}
\includegraphics[width=0.45\textwidth]{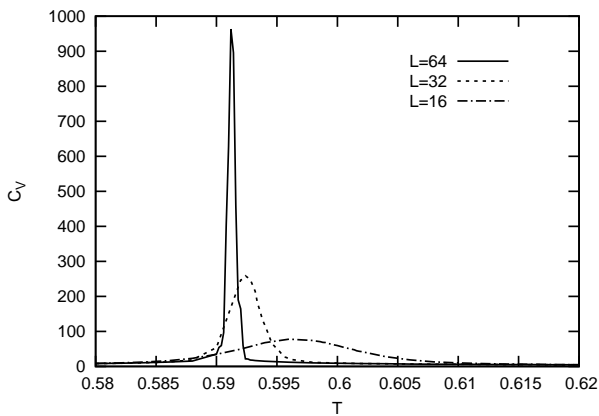}
\end{center}
\caption{Plot of the heat capacity as function of temperature in the case of the ten-state Potts model, for $q=0.95$ and different lattice sizes.}
\label{fig:2}
\end{figure}

In figure \ref{fig:2} we plotted the heat capacity as function of temperature for different lattice sizes and $q=0.95$. We can notice that the peak becomes sharper as the system size increases.

\begin{figure}[h]
\begin{center}
\includegraphics[width=0.45\textwidth]{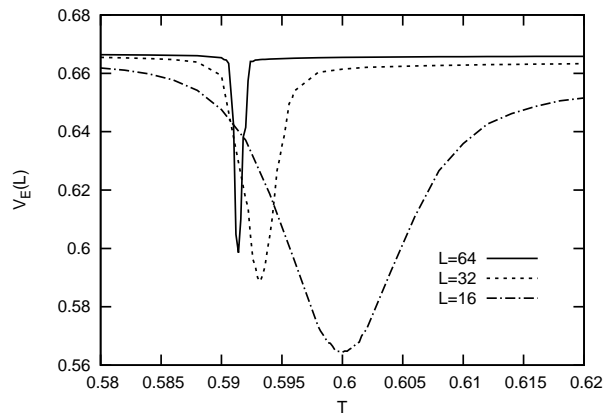}
\end{center}
\caption{Fourth order cumulant of the energy vs. temperature in the case of the ten-state Potts model, for $q=0.95$ and different lattice sizes.}
\label{fig:3}
\end{figure}

To evaluate the critical temperature of the transition we plotted the fourth order cumulant of the energy as function of temperature.
\begin{equation}\label{eq:16}
V_E(L) = 1 - \frac{ \langle E^4 \rangle_q}{3 \langle E^2 \rangle_q^2}
\end{equation}
It is known that $V_E(L)$ presents a minimum at the pseudo-critical temperature. Figure \ref{fig:3} shows the plot of $V_E(L)$ as function of temperature for $q=0.95$ and different lattice sizes.

\begin{figure}[h]
\begin{center}
\includegraphics[width=0.45\textwidth]{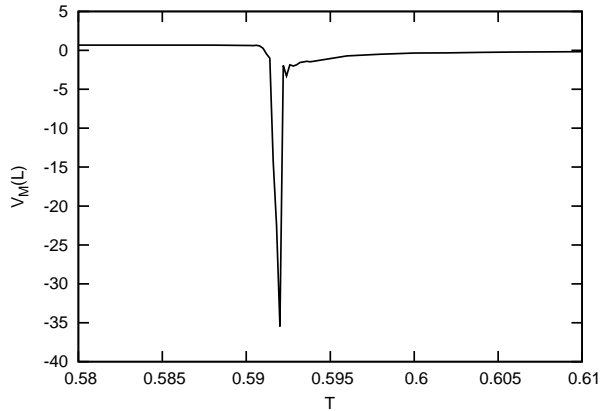}
\end{center}
\caption{Binder cumulant vs. temperature in the case of the ten-state Potts model, for $q=0.95$ and $L=64$.}
\label{fig:4}
\end{figure}

To establish the order of the phase transition we plotted the fourth order cumulant of the magnetization (the so-called Binder cumulant) as function of temperature (see Fig. \ref{fig:4}).
\begin{equation}\label{eq:17}
V_M(L) = 1 - \frac{ \langle M^4 \rangle_q}{3 \langle M^2 \rangle_q^2}
\end{equation}
The Binder cumulant exhibits a deep negative minimum at the pseudo-critical temperature, which indicates a first order phase transition \cite{tsai1998fourth}.

To evaluate the critical temperature in the thermodynamic limit, we plotted the pseudo-critical temperature as function of $1/L^2$, $L$ being the linear size of the lattice. The intercept of the linear fit gives us the critical temperature in the thermodynamic limit, i.e. for $L\to\infty$. In the case of the ten-state Potts model, for $q=0.95$ we obtained $T_c=0.5909(2)$.
\begin{figure}[h]
\begin{center}
\includegraphics[width=0.45\textwidth]{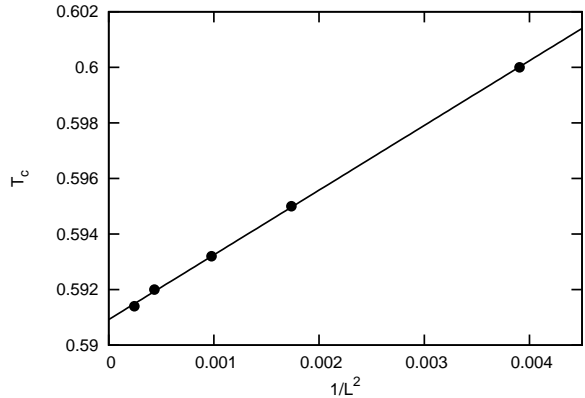}
\end{center}
\caption{Pseudo-critical temperature vs. $1/L^2$ in the case of the ten-state Potts model, for $q=0.95$.}
\label{fig:5}
\end{figure}

For the six-state Potts model the simulation results are similar, so we deal with a first order phase transition (for $q=1$, as well as $q\neq 1$).

\begin{figure}[h]
\begin{center}
\includegraphics[width=0.45\textwidth]{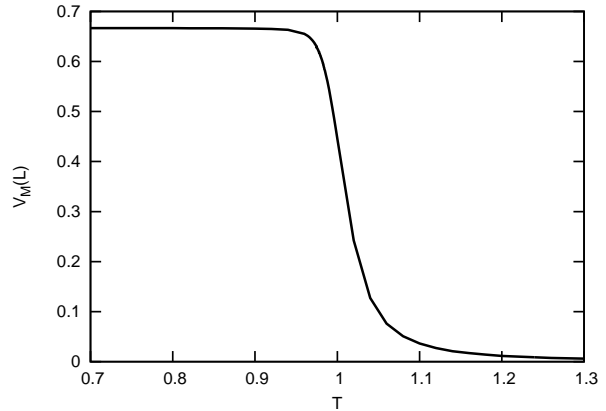}
\end{center}
\caption{Binder cumulant vs. temperature in the case of the two-state Potts model, for $q=0.95$ and $L=64$.}
\label{fig:6}
\end{figure}

The two-state Potts model presents a different behavior. The numerical results obtained by simulations show us that this model exhibits a second order phase transition. In Fig. \ref{fig:6} we plotted the Binder cumulant as function of temperature. The form of the plot indicates a continuous (second-order) phase transition.

To evaluate more accurately the critical temperature in the thermodynamic limit, we plotted the Binder cumulant as function of temperature for different lattice sizes. The curves intersect in one point, which corresponds to the critical temperature.
\begin{figure}[h]
\begin{center}
\includegraphics[width=0.45\textwidth]{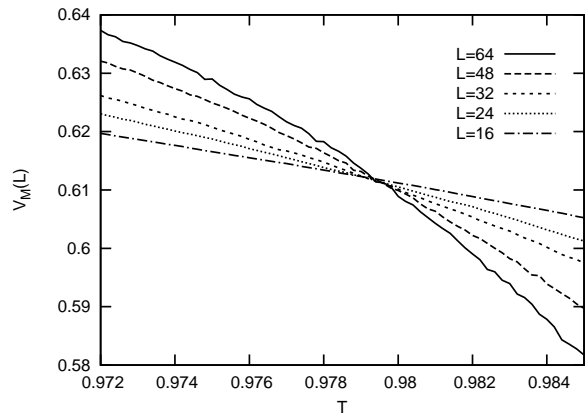}
\end{center}
\caption{Binder cumulant vs. temperature in the case of the two-state Potts model, for $q=0.95$ and different lattice sizes.}
\label{fig:7}
\end{figure}

\begin{table}[h]
\begin{ruledtabular}
\begin{tabular}{lllll}
{} & $q=1.00$ & $q=1.00$ & $q=0.95$ & $q=0.90$ \\
{} & (exact) & {} & {} & {} \\
$r=10$ & $0.701232$ & $0.7012(8)$ & $0.5909(2)$ & $0.4853(2)$ \\
$r=6$ & $0.807607$ & $0.808(1)$ & $0.6864(0)$ & $0.569(4)$ \\
$r=2$ & $1.134593$ & $1.134(2)$ & $0.979(4)$ & $0.828(4)$
\end{tabular}
\end{ruledtabular}
\caption{Critical temperature for different values of $r$ and $q$.}
\end{table}
The above table presents the numerical results obtained for the critical temperature, including all the previously discussed cases.
One can observe that the critical temperature has a strong dependence on the Tsallis entropic index $q$.

\section{Conclusions and final remarks} \label{sec:conclusions}
We performed Monte Carlo simulations of the standard Potts model using nonextensive statistics. The multiprocessor code used for the simulations was based on the Metropolis algorithm, were we replaced the classical Boltzmann-Gibbs probabilities with Tsallis acceptance ratios. Following the idea presented in Refs. \cite{crokidakis2009finite} and \cite{reis2002evidences} we used $1/\beta'$ as a temperature scale.

The results show that the two-dimensional Potts model undergoes a phase transition in the nonextensive case, i.e. for $q\neq 1$. The order of the phase transition was established from the behavior of the Binder cumulant. For the ten-state and six-state Potts model we deal with a first order phase transition, while for the two-state Potts model we have a second order transition. We evaluated the critical temperature in the thermodynamic limit for different values of the Tsallis entropic index and we found that it has a strong dependence on $q$.

\begin{acknowledgments}
I would like to thank my wife, Milena, for her understanding and support during the writing of this paper.
\end{acknowledgments}

\bibliography{ref_boer_2011.bib}

\end{document}